\documentclass[11pt]{article}
\long\def\comment#1{}

\usepackage{acsc}
\usepackage{epsfig}
\usepackage{graphicx}
\usepackage{xspace}
\usepackage{subfigure}
\usepackage{float}
\usepackage{amsmath}
\usepackage{amssymb}
\usepackage{latexsym}
\usepackage{algorithm}
\usepackage[noend]{algorithmic}
\usepackage{url}
\usepackage{mathrsfs}

\newcounter{example}[section]
\renewcommand{\theexample}{\nthesection.\arabic{example}}
\newenvironment{example}{
     \refstepcounter{example}
     {\vspace{1ex} \noindent\bf  Example  \theexample:}}{
     \eop\vspace{1ex}} 

\newcounter{definition}[section]
\renewcommand{\thedefinition}{\nthesection.\arabic{definition}}

\newcounter{theorem}[section]
\renewcommand{\thetheorem}{\thesection.\arabic{theorem}}
\newenvironment{theorem}{\begin{em}
        \refstepcounter{theorem}
        {\vspace{1ex} \noindent\bf  Theorem  \thetheorem:}}{
        \end{em}\eop\vspace{1ex}} 

\newcounter{lemma}[section]
\renewcommand{\thelemma}{\thesection.\arabic{lemma}}
\newenvironment{lemma}{\begin{em}
        \refstepcounter{lemma}
        {\vspace{1ex}\noindent\bf Lemma \thelemma:}}{
        \end{em}\eop\vspace{1ex}} 

\newcounter{remark}[section]
\renewcommand{\theremark}{\thesection.\arabic{remark}}

\newcommand{\proofsketch}{\noindent{\bf Proof Sketch: }}

\newcommand{\nthesection}{\arabic{section}}

\newcommand{\eop}{\hspace*{\fill}\mbox{$\Box$}}

\newcommand{\stitle}[1]{\vspace{1ex} \noindent{\bf #1}}

\newcommand{\X}{{\cal X}}

\newcommand{\ei}{\end{itemize}}
\newcommand{\ee}{\end{enumerate}}

\newcommand{\beqn}{\begin{eqnarray*}}
\newcommand{\eeqn}{\end{eqnarray*}}

\newcounter{ccc}

\newcommand{\eat}[1]{}

\newfloat{tcm}{thp}{loa}
\floatname{tcm}{Recursive \sql}

\newcommand{\sql}{{\sc sql}\xspace}

\newcommand{\ctab}{Table~}

\title{Fast Probabilistic Ranking under $x$-Relation Model}
\author{
Lijun Chang, Jeffrey Xu Yu and Lu Qin\\
The Chinese University of Hong Kong, Hong Kong, China\\
\{ljchang,yu,lqin\}@se.cuhk.edu.hk
}

\begin{document}
\maketitle

\begin{abstract}
The probabilistic top-$k$ queries based on the interplay of score
and probability, under the possible worlds semantic, become an
important research issue that considers both score and uncertainty
on the same basis.  In the literature, many different probabilistic
top-$k$ queries are proposed. Almost all of them need to compute the
probability of a tuple $t_i$ to be ranked at the $j$-th position
across the entire set of possible worlds. The cost of such computing is the
dominant cost and is known as $O(kn^2)$, where $n$ is the size of
dataset. In this paper, we propose a new novel algorithm that
computes such probability in $O(kn)$.
\end{abstract}

\section{Introduction}

Ranking is an import issue in uncertain data, and has attracted a lot
of attentions recently.
%
The probabilistic top-$k$ queries based on the interplay of score
and probability, under the possible worlds semantic, were first
studied in~\cite{ICDE07:Soliman}.
%
%
In this paper, we show that we can significantly improve the
performance for all the probabilistic top-$k$ queries in the
literature
\cite{ICDE08:Yi,TKDE08:Yi,ICDE08:Hua,SIGMOD08:Hua,VLDB08:Jin,DBRank08:Zhang,SSDBM08:Bernecker,EDBT08:Lian}
under the $x$-Relation model. We achieve it by proposing a new novel
algorithm to reduce the dominant cost of computing probabilistic
top-$k$ queries to be $O(kn)$, which is known to be $O(kn^2)$, where
$n$ is the size of the dataset.

\section{$x$-Relation Model and Probabilistic top-k semantics}

In the $x$-Relation model~\cite{VLDB06:Agrawal,TKDE08:Yi}, an
$x$-Relation contains a set of independent $x$-tuples (called
generation rules in~\cite{ICDE07:Soliman,SIGMOD08:Hua}). An
$x$-tuple consists of a set of mutually exclusive tuples (or called
alternatives) to represent a discrete probability distribution of
the possible tuples the $x$-tuple may take in a randomly
instantiated data.
In an $x$-tuple, each alternative $t$ has a score $score(t)$, and a
probability $p(t)$ that represents its existence probability over
possible instances.
%
%
In the $x$-Relation model, the alternatives of $x$-tuples are assumed
to be disjoint. In the following, we denote an $x$-Relation as $\X$,
an $x$-tuple as $\tau$, and call an alternative a tuple, denoted as
$t$.

\begin{example}
Fig.~\ref{fig:pdb:db} shows an $x$-Relation which consists of three
$x$-tuples, $\tau_1 = \{t_1, t_3\}$, $\tau_2 = \{t_2\}$, and $\tau_3 =
\{t_4\}$.  The $x$-tuple $\tau_1$ indicates a probability distribution
over $t_1$ and $t_3$, with probability $p(t_1) = 0.3$ for its true
content to be $t_1$, with probability $p(t_3) = 0.5$ for its true
content to be $t_3$, and with probability $1-p(t_1)-p(t_3) = 0.2$ for
none of $t_1$ and $t_3$ to be the true content.
\end{example}

\begin{figure}[t]
{
\begin{center}
    \subfigure[$x$-Relation] {
    \label{fig:pdb:db}
      \begin{tabular}[b]{|c|c|c|c|}
      \hline
      $x$-$tuple$ & $tuple$ & $score$  & $prob$ \\ \hline\hline
      $\tau_1$  & $t_1$  & $100$ & $0.3$ \\ \cline{2-4}
                & $t_3$  & $80$ & $0.5$ \\ \hline
      $\tau_2$  & $t_2$  & $90$ & $1.0$ \\ \hline
      $\tau_3$  & $t_4$  & $70$ & $0.8$ \\ \hline
      \end{tabular}
    }
    \subfigure[Possible Worlds] {
    \label{fig:pdb:pwd}
      \begin{tabular}[b]{|l|r||l|}
      \hline
      Possible world ($I$)  & $\Pr(I)$ & top-2 \\ \hline\hline
      $\{t_2\}$ & $(1-p(t_1)-p(t_3))p(t_2)(1-p(t_4)) = 0.04$ & $t_2$ \\ \hline
      $\{t_2,t_4\}$      & $(1-p(t_1)-p(t_3))p(t_2)p(t_4) = 0.16$ & $t_2, t_4$ \\ \hline
      $\{t_1,t_2\}$      & $p(t_1)p(t_2)(1-p(t_4)) = 0.06$ & $t_1, t_2$ \\ \hline
      $\{t_1,t_2,t_4\}$  & $p(t_1)p(t_2)p(t_4) = 0.24$ & $t_1, t_2$ \\ \hline
      $\{t_2,t_3\}$      & $p(t_3)p(t_2)(1-p(t_4)) = 0.10$ & $t_2, t_3$ \\ \hline
      $\{t_2,t_3,t_4\}$  & $p(t_3)p(t_2)p(t_4) = 0.40$ & $t_2, t_3$ \\ \hline
      \end{tabular}
    }
\end{center}
} \vspace*{-0.4cm} \caption{$x$-Relation Data} 
\label{fig:pdb}
\end{figure}

In general, an $x$-Relation, $\X$, is a probability distribution over
a set of possible instances $\{I_1, I_2, \cdots\}$. A possible
instance, $I_j$, maintains zero or one alternative for every $x$-tuple
$\tau \in \X$. The probability of an instance $I_j$, $\Pr(I_j)$, is
the probability that $x$-tuples take certain or none alternatives in
$I_j$, such that $\Pr(I_j) = \prod_{t \in I_j} p(t) \times \prod_{\tau
\notin I_j} (1-\Pr(\tau))$ where $\tau \notin I_j$ means $x$-tuple
$\tau$ takes no alternative in $I_j$ and $\Pr(\tau) = \sum_{t\in \tau}
p(t)$. The entire set of possible worlds of an $x$-Relation, $\X$, denoted as
$pwd(\X)$, is the set of all the subsets $I_j~( \subseteq \X)$ with
probability greater than 0 ($\Pr(I_j) > 0$).

\begin{example}
Fig.~\ref{fig:pdb:pwd} shows the total $6$ possible worlds for
the $x$-Relation in Fig.~\ref{fig:pdb:db}.
%
%
The possible world $\{t_1,t_2\}$ means that, $\tau_1$ takes the
alternative $t_1$, $\tau_2$ takes the alternative $t_2$, and
$\tau_3$ takes none. The probability of this possible world becomes
$p(t_1)p(t_2)(1-p(t_4)) = 0.06$.
%
%
Note that the sum of the probabilities of all the
possible worlds is equal to 1.
\end{example}

\stitle{Probabilistic top-k semantics}:
Several probabilistic top-$k$ semantics have been proposed recently
under the $x$-Relational model
including Uncertain Top-k Query (U-Topk)
\cite{ICDE07:Soliman,TKDE08:Yi}, Uncertain k-Ranks Query (U-kRanks)
\cite{ICDE07:Soliman,TKDE08:Yi}, Global-Topk
\cite{DBRank08:Zhang},
Probabilistic Threshold top-$k$ query (PT-k) \cite{SIGMOD08:Hua}, and
the Probabilistic $k$ top-$k$ query (Pk-topk) \cite{VLDB08:Jin}.
The PT-k and Pk-topk are similar to the Global-Topk.  Global Top-k
query finds $k$ tuples with the highest top-$k$ probability.  PT-k
finds all the tuples that have top-$k$ probability above a user-given
threshold. Pk-topk finds $k$ tuples with the highest top-$k$
probability
%
%
in a data stream environment, where every tuple is independent.
All the above existing solutions except U-Topk
%
%
need to compute the probability of a tuple, $t_i$, to be ranked at the
$j$-th position across the entire set of possible worlds, denoted
$p_{i,j}$.

Below, we introduce U-kRanks and Global Top-k with the emphasis on
how $p_{i,j}$ is used.
%
%
%
%
%
Let $p_{i,j}$ be the probability of a tuple $t_i$ to be ranked at
the $j$-th position across the entire set of possible worlds
\cite{ICDE07:Soliman,TKDE08:Yi}.
%
\begin{equation}
p_{i,j} = \sum_{I \in pwd(\X), t_i = \Psi_j(I)} \Pr(I)
\end{equation}
where $\Psi_j(I)$ denote the tuple with the $j$-th largest score in an
instance $I$ of the possible worlds.
The answer to a U-kRanks query on an $x$-Relation $\X$ is a vector
$(t_1^*,\cdots,t_k^*)$, where $t_j^* = \arg \max_{t_i} p_{i,j}$ for
$j = 1, \cdots, k$.
%
%
%
%
Let $tkp(t_i)$ be the top-$k$ probability of a tuple, $t_i$, which is
the marginal probability that $t_i$ is ranked top-$k$ in the possible
worlds \cite{DBRank08:Zhang}.
%
%
\begin{equation}
tkp(t_i) = \sum_{I \in pwd(\X), t_i \in topk(I)}\Pr(I) =
\sum_{j=1}^k p_{i,j} \label{eq:tkp}
\end{equation}
where $t_i \in topk(I)$ means that the tuple $t_i$ is ranked as one of the
top-$k$ tuples in the instance $I$.
The answer to a Global Top-k query on an $x$-Relation $\X$ is a set of
size $k$, $\{t_1^*,\cdots,t_k^*\}$, which satisfies $tkp(t_j^*) \geq
tkp(t)$ for any $j = 1,\cdots,k$ and $t \notin
\{t_1^*,\cdots,t_k^*\}$.

\begin{example}
The U-2Ranks query on Fig.~\ref{fig:pdb}(b) returns $2$ tuples, $t_2$
($score(t_2) = 90$) and $t_3$ ($score(t_3) = 80$), for $t_2$ is ranked
top and $t_3$ ranked 2nd.  The probability for $t_2$ to be ranked top
is $p_{2,1} = 0.04+0.16+0.1+0.4 = 0.7$ and the probability for $t_3$
to be ranked 2nd is $p_{3,2} = 0.1+0.4 = 0.5$.
The tuple $t_1$ has the highest score $100$ but with a low
probability $0.3$, therefore, it is not a result in U-2Ranks.
The Global Top-2 query returns a set of 2 tuples $\{t_2,t_3\}$. Here
$tkp(t_2) = 0.04 + 0.16 + 0.06 + 0.24 + 0.10 + 0.40 = 1.0$, because
$t_2$ is ranked as a top-$2$ tuple in every instance, and $tkp(t_3) =
0.10 + 0.40 = 0.5$, because $t_3$ is ranked as a top-$2$ tuple only in
two instances.
Note that the results of U-kRanks and Global Top-k do not
necessarily the same.
%
%
\end{example}

It is important to note that all these probabilistic ranking
queries, namely, U-kRanks, Global-Topk, PT-k, and Pk-topk, need to
compute the $p_{i,j}$ values for all $t_i \in \X$ and $j =
1,\cdots,k$, and computing $p_{i,j}$ is the dominant cost in such
probabilistic ranking queries.

\section{$p_{i,j}$ Computing}

We discuss $p_{i,j}$ computing for a given $k$ and an $x$-Relation $\X
= \{t_1,\cdots,t_n\}$ sorted in the descending score order.  For
simplicity and without loss of generality, in the following
discussions, we further assume there are no tie scores in $\X$ such
that $score(t_i) \neq score(t_j)$ for any $t_i \neq t_j$ in $\X$. Note
that all algorithms including our algorithm to be discussed can deal
with tie scores with minor modification for computing $p_{i,j}$.

\cite{TKDE08:Yi} showed that the time complexity of computing
$p_{i,j}$ for all $t_i \in \X$ and $j=1, \cdots, k$ is $O(kn^2)$.
We introduce it in brief below.

Given an $x$-Relation $\X = \{t_1,\cdots,t_n\}$ sorted in the
decreasing score order. Let $\X_i = \{t_1,\cdots,t_i\}$ denote a
reduced $x$-Relation on the largest $i$ tuples, together with the
projected (exclusive/independent) relationship between tuples.
It is obvious that $p_{i,j}$ is the same to be computed either on $\X$
or $\X_i$, under the $x$-Relation model.
Formally, let $\Pr(\tau | \X_i)$ be the existence probability of an
$x$-tuple $\tau$ with respect to $\X_i$ as follows.
\begin{equation}
\label{eq:probx} \Pr(\tau | \X_i) = \sum_{t \in \tau, t \in \X_i}
p(t)
\end{equation}
Then, $\Pr(\tau) = \Pr(\tau | \X)$.

We highlight the main idea of computing $p_{i,j}$ in $O(kn^2)$
\cite{TKDE08:Yi} below.
First, consider a special case, where every $x$-tuple contains only
one tuple (single-alternative), or equivalently, all the tuples are
independent.  Then, $p_{i,j}$ is equal to the probability that a
randomly generated possible world from $\X_i$ contains $t_i$ and
there are $j$ tuples in total. In other words, $p_{i,j}$ is the sum
of the probabilities of the possible worlds that contain $t_i$ and
there are exactly $j-1$ tuples taken from the set $\X_{i-1} =
\{t_1,\cdots,t_{i-1}\}$.
Let $r_{i,j}$ denote the probability that a randomly generated
possible world from $\X_i$ has exactly $j$ tuples, then $p_{i,j} =
p(t_i) \cdot r_{i-1,j-1}$. For the totally independent case,
%
%
the set of all $r_{i,j}$ values can be computed efficiently by the
following dynamic programming equation, in time complexity $O(kn)$.
\begin{equation}
r_{i,j} = \left\{\!
\begin{array}{ll}
p(t_i) \cdot r_{i-1,j-1} + (1-p(t_i)) \cdot r_{i-1,j}, &
  \text{if $i \geq j > 0$;} \\
(1-p(t_i)) \cdot r_{i-1,j}, & \text{if $i > j = 0$;} \\
1, & \text{if $i = j = 0$;}  \\
0, & \text{otherwise.} \label{eq:dp}
\end{array}\right.
\end{equation}
Second, consider the case where some $x$-tuples may contain multiple
tuples (multi-alternative). 
%
%
%
The noticeable difference is that $p_{i,j} \neq p(t_i) \cdot
r_{i-1,j-1}$ in the multi-alternative case, because an $x$-tuple
contains multiple-alternatives that are mutually exclusive. When it
needs to compute $p_{i,j}$ for a tuple $t_i$, the $x$-tuple that
contains $t_i$ may have other alternatives been computed already.
It needs to remember whether an alternative of an $x$-tuple has
already been computed in $\X_{i-1}$ using a set denoted ${\cal S}$.
Let ${\cal S} = \{\tau_1,\cdots,\tau_s\}$ be the set of $x$-tuples,
that have at least one alternative computed in $\X_{i-1}$ already,
with probability $\Pr(\tau_l | \X_{i-1})$ for $1 \leq l \leq s$ (Refer
to Eq.~(\ref{eq:probx})).
When $t_i$ appears and the $x$-tuple $\tau_x$ that contains $t_i$ has
already appeared in ${\cal S}$, it 
%
%
computes $p_{i,j}$ as $p_{i,j} = p(t_i) \cdot r_{s,j-1}^\prime$. Here,
$r_{i,j-1}^\prime$, for $1 \leq i \leq s$ and $1 \leq j \leq k$,
need to be recomputed based on ${\cal S} = \{\tau_1,\cdots,\tau_s\}$
with $\Pr(\tau_x | \X_{i-1}) = 0$ using Eq.~(\ref{eq:dp}), and takes
$O(s \cdot k)$ time.
In the worst case, it takes $O(i \cdot k)$ to compute $p_{i,j}$ for a
specific $i$. The time complexity to compute $p_{i,j}$ values, for $1
\leq i \leq n$ and $1 \leq j \leq k$, is $O(kn^2)$.

\begin{table}[h]
\begin{center}
      \begin{tabular}[b]{|c|c|}
      \hline
      $\tau_1$ & $\{t_1 (0.3), t_4 (0.4)\}$ \\ \hline
      $\tau_2$ & $\{t_2 (0.5), t_8 (0.2)\}$ \\ \hline
      $\tau_3$ & $\{t_3 (0.5), t_6 (0.5)\}$ \\ \hline
      $\tau_4$ & $\{t_5 (0.6), t_7 (0.3)\}$ \\ \hline
      \end{tabular}
\end{center}
\vspace*{-0.4cm}
\caption{Multi-alternative $x$-Relation}
\label{fig:xrelation}
\end{table}

\begin{example}
Consider an $x$-Relation, $\X$, in Table~\ref{fig:xrelation} with 4
$x$-tuples, $\{\tau_1, \tau_2, \tau_3, \tau_4\}$ and 8 tuples
$\{t_1,\cdots,t_8\}$. Each $x$-tuple contains two tuples (alternatives).
We assume $score(t_i) > score(t_j)$ if $i < j$, and give the
probability of each tuple $t_i$, $p(t_i)$, in the corresponding
parentheses. For example, $\tau_1$ has two tuples $t_1$ and $t_4$
where $p(t_1) = 0.3$ and $p(t_4) = 0.4$.
Let $k = 2$. We show how to compute $p_{i,j}$ for all tuples $t_i$,
for $1 \leq i \leq 8$ and $j = 1, 2$.

Let all 8 tuples in $\X$ be sorted in the decreasing score order,
%
%
and let ${\cal S}$ be the set of $x$-tuples that have
multi-alternatives in $\X_{i-1}$.
Initially, $\X_0 = \emptyset$, ${\cal S} = \emptyset$.

First, consider $t_1$ which is the tuple that has the largest score,
and $\mathcal{S} = \emptyset$ implies that $t_1$ has no preceding
alternatives.  Because $r_{0,0}^\prime = 1$ and $r_{0,1}^\prime =
0$, thus $p_{1,1} = p(t_1) \cdot r_{0,0}^\prime = 0.3$ and $p_{1,2}
= p(t_1) \cdot r_{0,1}^\prime = 0$. $\X_1 = \{t_1\}$. Based on
Eq.~(\ref{eq:probx}), the current existence probability of $\tau_1$
in $\X_1$ is $\Pr(\tau_1 | \X_1) = p(t_1) = 0.3$.
${\cal S}$ is updated to be $\mathcal{S} = \{\tau_1\}$, because the
$x$-tuple $\tau_1$ contains $t_1$ that has been computed.  For
simplicity, we use $\mathcal{S} = \{\tau_1(0.3)\}$ to indicate that
${\cal S}$ contains $\tau_1$ whose current existence probability is
$0.3$.

Second, consider the second largest score tuple $t_2$, which has no
preceding alternatives computed, because the $x$-tuple $\tau_2$ that
contains $t_2$ does not appear in $\mathcal{S} = \{\tau_1 (0.3)\}$.
Because $r_{1,0}^\prime = 0.7$ and $r_{1,1}^\prime = 0.3$, thus
$p_{2,1} = p(t_2) \cdot r_{1,0}^\prime = 0.35$ and $p_{2,2} = 0.15$.
$\X_2 = \{t_1, t_2\}$. Based on Eq.~(\ref{eq:probx}), the current
existence probability of $\tau_2$ in $\X_2$ is $\Pr(\tau_2 | \X_2) =
p(t_2) = 0.5$. ${\cal S} = \{\tau_1 (0.3), \tau_2 (0.5)\}$.

In a similar fashion, the third largest score tuple $t_3$ is
computed which has no preceding alternatives in ${\cal S}$.  Because
$r_{2,0}^\prime = 0.35$ and $r_{2,1}^\prime = 0.5$, thus $p_{3,1} =
0.5 \cdot 0.35 = 0.175$ and $p_{3,2} = 0.25$. $\X_3 = \{t_1, t_2,
t_3\}$. Based on Eq.~(\ref{eq:probx}), the current existence
probability of $\tau_3$ in $\X_3$ is $\Pr(\tau_3 | \X_3) = p(t_3) =
0.5$. $\mathcal{S} = \{\tau_1 (0.3), \tau_2 (0.5), \tau_3
(0.5)\}$.

Fourth, consider the fourth largest score tuple $t_4$.
Note that the current $\mathcal{S} =
\{\tau_1 (0.3), \tau_2 (0.5), \tau_3 (0.5)\}$.  But because tuple
$t_4$ has a preceding alternative $t_1$ in $x$-tuple $\tau_1$ which
appears in ${\cal S}$ already, the existence probability of
$\Pr(\tau_1 | \X_3) = 0$ is reset. Therefore, ${\cal S}$ is updated
to be $\mathcal{S} = \{\tau_1 (0), \tau_2 (0.5), \tau_3 (0.5)\}$.
In order to compute $r_{3,0}^\prime$ and $r_{3,1}^\prime$, all the
$r_{i,j}^\prime$ values, for $i = 1,2$ and $j = 0,1$, need to be
recomputed as well based on the updated $\mathcal{S}$. Because
$r_{1,0}^\prime = 1$, $r_{1,1}^\prime = 0$, $r_{2,0}^\prime = 0.5$,
$r_{2,1}^\prime = 0.5$, $r_{3,0}^\prime = 0.25$, and $r_{3,1}^\prime =
0.5$, thus $p_{4,1} = p(t_4) \cdot r_{3,0}^\prime = 0.1$ and $p_{4,2}
= 0.2$.
$\X_4 = \{t_1, t_2, t_3, t_4\}$. Based on Eq.~(\ref{eq:probx}), the
current existence probability of $\tau_1$ in $\X_4$ is $\Pr(\tau_1 |
\X_4) = p(t_1) + p(t_4) = 0.3 + 0.4 = 0.7$. Therefore, $\mathcal{S} =
\{\tau_1 (0.7), \tau_2 (0.5), \tau_3 (0.5)\}$, which will be used in
the next iteration.
%

The same procedure repeats until all $p_{i,j}$ for all $t_i \in \X$
and $j = 1, 2$ are computed.
\end{example}

Note that, between consecutive computations of $p_{i,j}$ and
$p_{i+1,j}$, some $r_{s,j}^\prime$ computing cost can be shared
\cite{SIGMOD08:Hua,TKDE08:Yi}. \cite{SIGMOD08:Hua} also studied
several heuristics to fast compute $p_{i,j}$ but in the worst case it
is $O(kn^2)$.

\section{A New Novel Algorithm}
\label{sec:tkp}

In this paper, we propose a novel $O(kn)$ algorithm using a newly
introduced conditional probability $c_{i,j}$ given below,
\begin{equation}
c_{i,j} = \Pr(\text{Exactly $j$ tuples  appear in
$\{t_1,\cdots,t_i\}$ $|$ $t_{i+1}$ appears})
\label{eq:c}
\end{equation}
to fast compute $p_{i,j}$. Consider a general multi-alternative
case. Let $\X_i = \{t_1, \cdots, t_i\}$ be the set computed already.
Now, we consider $t_{i+1}$, assume $t_{i+1}$ appears. Among the
tuples computed already in $\X_i$, there may exist several tuples in
$\X_i$ that are contained in the same $x$-tuple that contains
$t_{i+1}$. Those tuples need to be removed in order to compute for
$t_{i+1}$, as we discussed in the previous section by setting the
existence probability to be zero.  Eq.~(\ref{eq:c}) is the
conditional probability of having exactly $j$ tuples in $\X_i =
\{t_1, \cdots, t_i\}$ after removing those tuples in $\X_i$ that are
contained in the same $x$-tuple that contains $t_{i+1}$, given
$t_{i+1}$ appears. It is interesting to note that
\begin{eqnarray}
p_{i,j} &=& \Pr(t_i \text{ appears}) \cdot \Pr(\text{Exactly $j$-$1$
tuples appear in
$\{t_1,\cdots,t_{i-1}\}$ $|$ $t_i$ appears}) \nonumber \\
&=& p(t_i) \cdot c_{i-1,j-1}
\label{eq:newpij}
\end{eqnarray}
%
%
And the problem becomes how to compute $c_{i,j}$ efficiently. Note
that there is no obvious relationship between $c_{i,j}$ and
$c_{i-1,j}$ (refer to Eq.~(\ref{eq:dp})). However, we observe that
there is a similar relationship between $c_{i,j}$ and $r_{i,j}$. Let
$\tau_x$ be the $x$-tuple that contains $t_{i+1}$. Then, the
relationship between $c_{i,j}$ and $r_{i,j}$ becomes as follows,
\begin{equation}
r_{i,j} = \left\{\!
\begin{array}{ll}
(1 - \Pr(\tau_x | \X_i)) \cdot c_{i,j}, & \text{if $j = 0$;} \\
(1 - \Pr(\tau_x | \X_i)) \cdot c_{i,j} + \Pr(\tau_x | \X_i) \cdot
c_{i,j-1}, &
  \text{if $j > 0$;} \label{eq:rc}
\end{array}\right.
\end{equation}

\begin{lemma}
Eq.~(\ref{eq:rc}) correctly computes $r_{i,j}$, given $c_{i,j}$.
\end{lemma}

\proofsketch
Assume that $c_{i,j}$ for $0 \leq j \leq k-1$ are correct as defined,
the probability that a randomly generated possible world has exactly
$j$ tuples from $\X_i$ is conditioned by the appearance of
$t_{i+1}$. Let $\tau_x$ be the $x$-tuple that has $t_{i+1}$, and $\rho$
denote $\Pr(\tau_x | \X_i)$. There are two cases.

First, $t_{i+1}$ has no preceding alternative, equivalently $\rho =
0$.  Then the two parts in the conditional probability $c_{i,j}$ are
independent, $c_{i,j} = \Pr($Exactly $j$ tuples appear in
$\{t_1,\cdots,t_i\}$), where the latter part of the equation is
actually $r_{i,j}$. Hence, Eq.~(\ref{eq:rc}) correctly computes
$r_{i,j}$, given that $c_{i,j}$ are correct.

Second, $t_{i+1}$ has some preceding alternatives, equivalently $\rho
> 0$. Assume that ${\cal S} = \{\tau_1,\cdots,\tau_s,\tau_x\}$ is the
set of $x$-tuples that have alternatives appearing in $\X_i =
\{t_1,\cdots,t_i\}$, where $\Pr(\tau_l | \X_i) > 0$ for all $\tau_l
\in {\cal S}$. Then $c_{i,j}$ is the probability that a randomly
generated possible world from $\{\tau_1,\cdots,\tau_s\}$ ($= {\cal S}
\setminus \{\tau_x\}$) has exactly $j$ $x$-tuples, and $r_{i,j}$ is the
probability that a randomly generated possible world from
$\{\tau_1,\cdots,\tau_s,\tau_x\}$ has exactly $j$ $x$-tuples. Hence,
Eq.~(\ref{eq:rc}) is correct based on the same idea shown in
Eq.~(\ref{eq:dp}). \eop

Given $c_{i,j}$ we can compute $r_{i,j}$ using Eq.~(\ref{eq:rc}).
The reverse also holds such that, given $r_{i,j}$, we can compute
$c_{i,j}$ correctly by the system of linear equations defined in
Eq.~(\ref{eq:rc}).
A general system of linear equations with $n$ equations and $n$
variables needs time $O(n^3)$. But the system of linear equations
defined by Eq.~(\ref{eq:rc}) has a special form, there are only two
diagonals of the coefficient matrix which are non-zero, so it can be
solved in $O(n)$ time \cite{lay02}. In our problem, there are $k$
linear equations with $k$ variables, it can be solved in time $O(k)$,
using $c_{i,0} = r_{i,0}/(1- \rho)$ and $c_{i,j} = (r_{i,j} - \rho
\cdot c_{i,j-1})/(1-\rho)$ where $\rho = \Pr(\tau_x | \X_i)$, for $1
\leq j \leq k-1$. Note that $0 < \Pr(\tau_x | \X_i) < 1$.
In addition, given $c_{i,j}$, $r_{i+1,j}$ can also be computed using
Eq.~(\ref{eq:rc}), by replacing $\Pr(\tau_x | \X_i)$ with $\Pr(\tau_x
| \X_{i+1})$, where $\tau_x$ is the $x$-tuple that contains $t_{i+1}$.

The algorithm to compute $r_{i,j}$ and $p_{i,j}$ values for a tuple
$t_i$ is shown in Algorithm~\ref{alg:prob}. It takes three inputs,
namely, the tuple $t_i$, the $r_{i-1,j}$ values, and a set of
$x$-tuples, ${\cal S} = \{\tau_1, \cdots, \tau_s\}$, that have been
computed with their probability $\Pr(\tau_l) = \Pr(\tau_l |
\X_{i-1})$. It first computes $\rho$ (line 1-2). Then, it computes
the $c_{i-1,j}$ values by solving a system of linear equations
defined by Eq.~(\ref{eq:rc}) (line 3-5), and computes the $p_{i,j}$
values (line 6). In line 7-10, it computes the $r_{i,j}$ values
using Eq.~(\ref{eq:rc}).  Finally, it updates the probability
$\Pr(\tau_x)$ (line~11-14). Note that, in our algorithm, the only
values needed to compute $p_{i,j}$ values are $r_{i-1,j}$ values and
$\Pr(\tau_x | \X_{i-1})$.

\begin{algorithm}[t]
\caption{CondProb(${\cal S}$, ${\cal R}_{i-1}$,$t_i$)}
\label{alg:prob}
 {
\begin{tabbing}
{\bf\ Input}: \hspace{0.3cm}\= the probability for $x$-tuples
${\cal S} = \{\tau_1(\Pr(\tau_1)),\cdots,\tau_s(\Pr(\tau_s))\}$ \\
  \>${\cal R}_{i-1} = \{r_{i-1,0},\cdots,r_{i-1,k-1}\}$ and a tuple $t_i$.\\
{\bf\ Output}: \> $r_{i,j-1}$ and $p_{i,j}$, for $1 \leq j \leq k$.
\end{tabbing}

 \begin{algorithmic}[1]
 \STATE Let $\tau_x$ be the $x$-tuple that has $t_i$;
 \STATE $\rho \leftarrow \Pr(\tau_x)$ if $\tau_x (\Pr(\tau_x))$
        appears in ${\cal S}$ otherwise $0$; \\
        // compute $c_{i-1,j}$ and $p_{i,j}$ for $0 \leq j \leq k-1$
 \STATE $c_{i-1,0} \leftarrow r_{i-1,0}/(1-\rho)$;
 \FOR{ $j \leftarrow 1$ \textbf{to} $k-1$ }
    \STATE $c_{i-1,j} \leftarrow (r_{i-1,j} - \rho \cdot c_{i-1,j-1})
 / (1 - \rho)$;
 \ENDFOR
 \STATE $p_{i,j} \leftarrow p(t_i) \cdot c_{i-1,j-1}$, for $1 \leq j \leq
 k$; \\
        // compute $r_{i,j}$ for $0 \leq j \leq k-1$
 \STATE $\rho \leftarrow \rho + p(t_i)$;
 \STATE $r_{i,0} \leftarrow (1-\rho) \cdot c_{i-1,0}$;
 \FOR{ $j \leftarrow 1$ \textbf{to} $k-1$ }
    \STATE $r_{i,j} \leftarrow (1-\rho) \cdot c_{i-1,j} + \rho \cdot
    c_{i-1,j-1}$;
 \ENDFOR
 \IF {$\tau_x \not \in {\cal S}$}
 \STATE ${\cal S} \leftarrow {\cal S} \cup \{\tau_x(\rho)\}$;
 \ELSE
 \STATE update ${\cal S}$ by changing $\Pr(\tau_x)$ to be $\rho$;
 \ENDIF
 \STATE {\bf return} (${\cal S}$, $\{r_{i,0},\cdots,r_{i,k-1}\}$,
 $\{p_{i,1},\cdots,p_{i,k}\})$;
 \end{algorithmic}
 }
\end{algorithm}

\begin{theorem}
\label{the:prob} Algorithm~\ref{alg:prob} correctly computes the
$p_{i,j}$ values with time complexity of $O(k)$.
\end{theorem}

\proofsketch It is obvious from the discussions above. \eop

In order to compute all $p_{i,j}$, we enumerate all $t_i \in \X$,
which is sorted in the descending order score, such as $score(t_i) >
score(t_j)$ if $i < j$ as given below.
%
%
\begin{algorithmic}[1]
\STATE Let ${\cal S} = \emptyset$; \STATE Let ${\cal R}_0 =
\{r_{0,0}, r_{0,1}, \cdots, r_{0,k-1}\}$ where
       $r_{0,j}$, for $0 \leq j \leq k-1$, are computed;
\FOR {$i = 1$ to $n$}
\STATE $({\cal S}, {\cal R}_i, {\cal P}_i) \leftarrow$ CondProb(${\cal
  S}$, ${\cal R}_{i-1}$, $t_i$);
\STATE output ${\cal P}_i = \{p_{i,1}, p_{i,2}, \cdots, p_{i,k}\}$;
\ENDFOR
\end{algorithmic}
It is obvious that the time complexity to compute all $p_{i,j}$ is
$O(kn)$.

\begin{figure}[t]
\begin{center}
    \subfigure[The Existing $O(kn^2)$ Approach] {
        \includegraphics[scale=0.8]{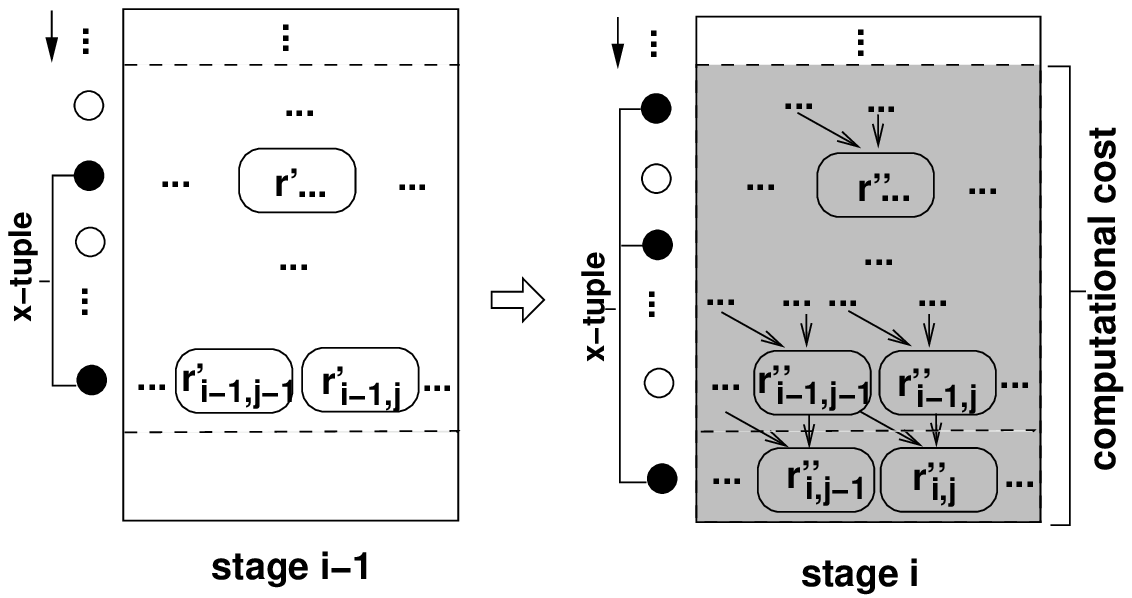}
    }
    \subfigure[Our New $O(kn)$ Approach] {
        \includegraphics[scale=0.8]{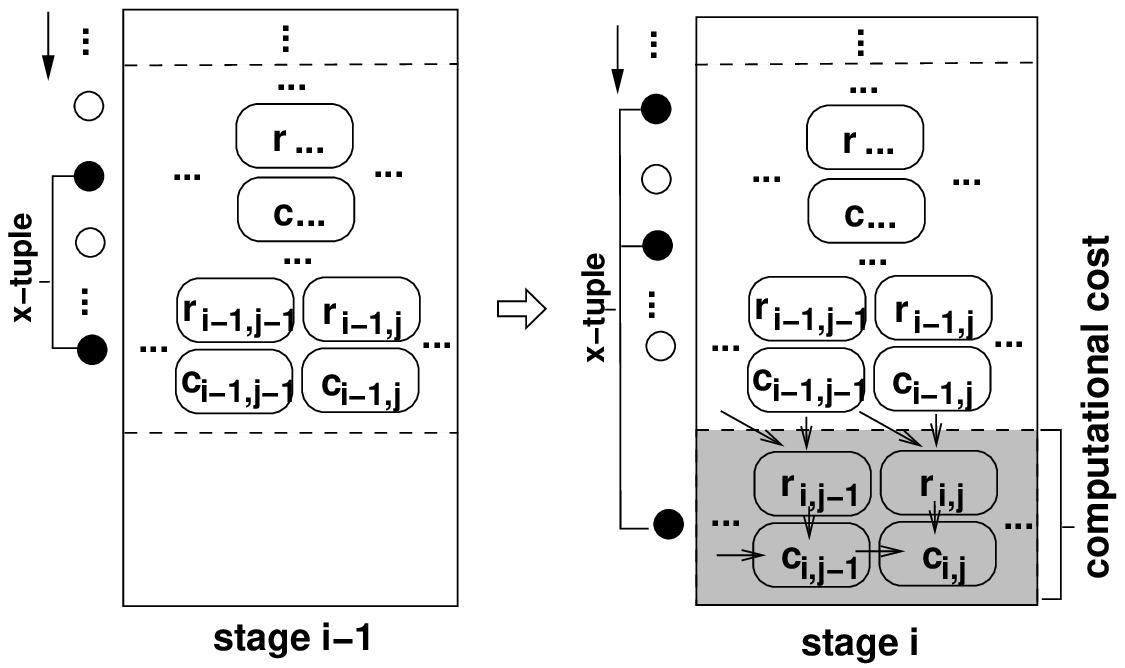}
    }
\end{center}
\vspace*{-0.4cm}
\caption{Computational Cost}
\vspace*{-0.4cm}
\label{fig:cost}
\end{figure}

Fig.~\ref{fig:cost}(a) illustrates the existing $O(kn^2)$ approach
to compute $r_{i,j}^\prime$ in the stage $i$ based on the stage
$i$-$1$. Note that the stage $i$ is the $i$-iteration to compute for
the $i$-th largest score tuple in $\X_i$. On the left side in the
stage $i$-$1$ and the stage $i$, it indicates that some $x$-tuple
(marked by $\bullet$) contains several tuples (alternatives).
On the other hand, Fig.~\ref{fig:cost}(b) illustrates our $O(kn)$
approach to compute $r_{i,j}$, using $c_{i,j}$, in the stage $i$
based on the stage $i$-$1$. The shaded parts in
Fig.~\ref{fig:cost}(a)(b) indicate the equations needed to compute,
and the difference between the two shaded regions confirms the
significant cost saving of our approach.

\begin{example}
Consider the example $x$-Relation in Table~\ref{fig:xrelation}. We
show the steps of our algorithm to compute $p_{i,j}$.  Let $k = 2$.
We denote the sequence of $x$-tuples that have been scanned as
$\mathcal{S}$.
Initially, $\X_0 = \emptyset$, ${\cal S} = \emptyset$, $r_{0,0} = 1$
and $r_{0,1} = 0$.

First, consider $t_1$ which is the largest score tuple. It has no
preceding alternatives, $\Pr(\tau_1) = 0$, $c_{0,0} = 1$ and
$c_{0,1} = 0$. Then, $p_{1,1} = p(t_1) \cdot c_{0,0} = 0.3$ and
$p_{1,2} = 0$.
After computing $t_1$, $\X_1 = \{t_1\}$, $\mathcal{S} = \{\tau_1
(0.3)\}$, and we have $r_{1,0} = (1-\Pr(\tau_1)) \cdot c_{0,0} =
0.7$ and $r_{1,1} = (1-\Pr(\tau_1)) \cdot c_{0,1} + \Pr(\tau_1)
\cdot c_{0,0} = 0.3$.

The second largest score tuple $t_2$
has no preceding alternatives, $\Pr(\tau_2) = 0$, $c_{1,0} = r_{1,0} =
0.7$ and $c_{1,1} = 0.3$.  Then, $p_{2,1} = p(t_2) \cdot c_{1,0} =
0.35$ and $p_{2,2} = 0.15$.
After computing $t_2$, $\X_2 = \{t_1, t_2\}$, $\mathcal{S} =
\{\tau_1 (0.3), \tau_2 (0.5)\}$, and in addition we have $r_{2,0} =
0.35$ and $r_{2,1} = 0.5$.

The third largest score tuple
$t_3$ has no preceding alternatives $\Pr(\tau_3) = 0$, $c_{2,0} =
0.35$ and $c_{2,1} = 0.5$. Then, $p_{3,1} = p(t_3) \cdot c_{2,0} =
0.175$ and $p_{3,2} = 0.25$.
After computing $t_3$, $\X_3 = \{t_1, t_2, t_3\}$, $\mathcal{S} =
\{\tau_1 (0.3), \tau_2 (0.5), \tau_3 (0.5)\}$, and in addition we
have $r_{3,0} = 0.175$ and $r_{3,1} = 0.425$.

The fourth largest score tuple $t_4$ has a preceding alternative
$t_1$ that is contained in $x$-tuple $\tau_1$ which appears in
${\cal S}$. Therefore, $\rho = \Pr(\tau_1 | \X_3) = 0.3$, $c_{3,0} =
r_{3,0}/(1- \rho) = 0.25$, $c_{3,1} = (r_{3,1} - \rho\cdot
c_{3,0})/(1-\rho) = 0.5$, $p_{4,1} = p(t_4) \cdot c_{3,0} = 0.1$ and
$p_{4,2} = 0.2$.  After computing $t_4$, $\X_4 = \{t_1, t_2, t_3,
t_4\}$, $\mathcal{S} = \{\tau_1 (0.7), \tau_2 (0.5), \tau_3
(0.5)\}$, and in addition we have $r_{4,0} = 0.075$ and $r_{4,1} =
0.325$.

The same procedure repeats until all $p_{i,j}$ for all $t_i \in \X$
and $j = 1, 2$ are computed.
\comment{
For $t_5$, $\Pr(\tau_4) = 0$, $c_{4,0} = 0.075$ and $c_{4,1} =
0.325$, $p_{5,1} = p(t_5) \cdot c_{4,0} = 0.045$ and $p_{5,2} =
0.195$. Update with $t_5$, $\mathcal{S} = \{\tau_1 (0.7), \tau_2
(0.5), \tau_3 (0.5), \tau_4 (0.6) \}$, $r_{5,0} = 0.03$ and $r_{5,1}
= 0.175$.
}
\end{example}

\section{Top-k Generator}
\label{sec:topk}

Algorithm~\ref{alg:prob} returns the set of $p_{i,j}$, which can be
used to compute the top-$k$ probability of a tuple, e.g. $tkp(t_i) =
\sum_{j=1}^k p_{i,j}$. A naive way to get the top-$k$ result is to
first compute the top-$k$ probabilities for all tuples, then report
the top-$k$ tuples with respect to the top-$k$ probability. In the
following, we will first discuss an upper bound, and then propose an
early stop condition, which avoids to retrieve all the tuples.

\begin{lemma}
Let $\{t_1,\cdots,t_i,\cdots\}$ be the order we scan the tuples, or
equivalently it is the decreasing score order, and $r_{i,j}$ is
defined as above. Then $tkp(t_{i+1}) \leq \sum_{j=1}^k r_{i,j}$, for
all $i \geq 1$. This upper bound is also tight for an arbitrary
sequence of tuples.
\end{lemma}

\proofsketch
Let $\tau_x$ be the $x$-tuple that have $t_{i+1}$, and $p = \Pr(\tau_x
| \X_i)$. Note that $p$ may be zero, or equivalently $t_{i+1}$ has
no preceding alternative. By Eq.~(\ref{eq:rc}), sum up the
$r_{i,j}$'s, $\sum_{j=0}^{k-1} r_{i,j} = \sum_{j=0}^{k-1} c_{i,j} -
p \cdot c_{i,k-1}$. We have
\begin{equation*}
\begin{array}{rcl}
\smallskip
tkp(t_{i+1}) & = & \sum_{j=1}^k p_{i+1,j} \\
\smallskip
& = &
p(t_{i+1}) \cdot \sum_{j=0}^{k-1} c_{i,j} \\
\smallskip
& \leq &
(1-p) \cdot \sum_{j=0}^{k-1} c_{i,j} \\
\smallskip
& = &
\sum_{j=0}^{k-1} c_{i,j} - p \cdot \sum_{j=0}^{k-1} c_{i,j} \\
\smallskip
& \leq & \sum_{j=0}^{k-1} c_{i,j} - p \cdot c_{i,k-1} \\
\smallskip
&
= & \sum_{j=0}^{k-1} r_{i,j}
\end{array}
\end{equation*}
where the third inequality holds because $\Pr(\tau_x | \X_i) +
p(t_{i+1}) \leq 1$, as $t_{i+1}$ is an alternative of $x$-tuple
$\tau_x$. So $tkp(t_{i+1}) \leq \sum_{j=1}^k r_{i,j}$. When
$p(t_{i+1}) = 1$, $\Pr(\tau_x | \X_i) = 0$, the above inequalities
hold with equality, and therefore $tkp(t_{i+1}) = \sum_{j=1}^k
r_{i,j}$. Hence this upper bound is tight.  \eop

\begin{lemma}
$\sum_{j=0}^{k-1} r_{i,j}$ are in decreasing order, e.g.
$\sum_{j=0}^{k-1} r_{i,j} \geq \sum_{j=0}^{k-1} r_{i+1,j}$, for any
$i \geq 1$.
\end{lemma}

\proofsketch
There are two cases, $t_{i+1}$ has preceding alternatives or not.

First, if $t_{i+1}$ does not have preceding alternatives, then
$r_{i+1,j}$ can be computed by Eq.~(\ref{eq:dp}). Summing up
$r_{i+1,j}$, we have $\sum_{j = 0}^{k-1} r_{i+1,j} = \sum_{j =
0}^{k-1} r_{i,j} - p(t_{i+1}) r_{i,k-1} \leq \sum_{j = 0}^{k-1}
r_{i,j}$.
Second, if $t_{i+1}$ has preceding alternatives, assuming $t_{i+1}$
is in the $x$-tuple $\tau_x$, then $\Pr(\tau_x | \X_i) > 0$. Assume
that $\{\tau_1,\cdots,\tau_x,\cdots,\tau_s\}$ is the set of $x$-tuples
that have alternatives in $\{t_1,\cdots,t_i\}$, with probability
$\Pr(\tau | \X_i)$. Then $\{\tau_1,\cdots,\tau_x,\cdots,\tau_s\}$ is
also the set of $x$-tuples that have alternatives in
$\{t_1,\cdots,t_i,t_{i+1}\}$, and their probability is $\Pr(\tau |
\X_{i+1})$, with $\Pr(\tau | \X_{i+1}) = \Pr(\tau | \X_i)$ for all
$x$-tuple $\tau$ except $\tau_x$, which has $\Pr(\tau_x | \X_{i+1}) >
\Pr(\tau_x | \X_i)$.
Let $r_{*,j}$ be the probability that a random generated possible
world from $\{\tau_1,\cdots,\tau_x,\cdots,\tau_s\}/\tau_x$, with
probabilities $\Pr(\tau | \X_i)$, has exactly $j$ $x$-tuples. The
relationship between $r_{i,j}$ and $r_{*,j}$, or between $r_{i+1,j}$
and $r_{*,j}$, is the same as Eq.~(\ref{eq:dp}) or
Eq.~(\ref{eq:rc}). Then $\sum_{j=0}^{k-1} r_{i,j} = \sum_{j=0}^{k-1}
r_{*,j} - \Pr(\tau_x | \X_i) \cdot r_{*,k-1}$, and $\sum_{j=0}^{k-1}
r_{i+1,j} = \sum_{j=0}^{k-1} r_{*,j} - \Pr(\tau_x | \X_{i+1}) \cdot
r_{*,k-1}$. So $\sum_{j=0}^{k-1} r_{i+1,j} \leq \sum_{j=0}^{k-1}
r_{i,j}$, as $\Pr(\tau_x | \X_{i+1}) > \Pr(\tau_x | \X_i)$. \eop

\begin{theorem}
\label{the:stop} If all the top-$k$ probabilities of the current
top-$k$ result, e.g. from the set $\{t_1,\cdots,t_i\}$, are greater
than or equal to $\sum_{j=0}^{k-1} r_{i,j}$, then we can stop, and
guaranty that any potential results in $\{t_{i+1}, \cdots, t_N\}$
can not be in the top-$k$ result.
\end{theorem}

With Theorem~\ref{the:stop}, we can develop an algorithm to compute
the top-$k$ tuples with respect to their top-$k$ probabilities,
which is shown in Algorithm~\ref{alg:topk}. It initializes in line
1-5, and $upBound$ denotes the upper bound of the top-$k$
probabilities of the remaining tuples (line~5). While the stop
condition is not satisfied (line 6), it retrieves the next largest
score tuple (line 7), computes its top-$k$ probability, inserts it
into the top-$k$ set (line 8-10), and update the upper bound (line
11). The top-$k$ set is maintained as a min-heap with size of $k$,
$top\text{-}k[k].tkp$ (line 6) is the minimum top-$k$ probability in
the min-heap. When inserting a new tuple associate with its top-k
probability into min-heap, if its top-$k$ probability is smaller
than that at the top of the min-heap, we do not need to insert it.
Otherwise, we replace the top tuple of the min-heap with the new
tuple and update the heap structure.

\begin{algorithm}[t]
\caption{ Top-k (k)} \label{alg:topk}
 {\small
\begin{tabbing}
{\bf\ Input}: \hspace{0.3cm}\= an integer k, specify the top-$k$ value,\\
{\bf\ Output}: \> top-$k$ tuples.
\end{tabbing}

 \begin{algorithmic}[1]
 \STATE Let $\{\tau_1,\cdots,\tau_m\}$ be the set of all the $x$-tuples;
 \STATE Initialize $\Pr(\tau_i) \leftarrow 0$, for $1 \leq i \leq
 m$;
 \STATE $top\text{-}k \leftarrow \emptyset$;
 \STATE Initialize $r_0 = 1$ and $r_j = 0$ for $1 \leq j \leq k-1$;
 \STATE $upBound \leftarrow \sum_{j=0}^{k-1} r_j$;
 \WHILE{ $top\text{-}k[k].tkp < upBound$ }
    \STATE $t \leftarrow Next()$;
    \STATE $r_j, p_j \leftarrow Prob(\{\Pr(\tau_1), \cdots, \Pr(\tau_m)\}, \{r_0,\cdots,r_{k-1}\}, t)$;
    \STATE $tkp(t) \leftarrow \sum_{j=1}^k p_j$;
    \STATE Insert $t$ into $top\text{-}k$;
    \STATE $upBound \leftarrow \sum_{j=0}^{k-1} r_j$;
 \ENDWHILE
 \STATE {\bf return} $top\text{-}k$;
 \end{algorithmic}
 }
\end{algorithm}

\begin{theorem}
Algorithm~\ref{alg:topk} correctly returns the top-$k$ tuples with
highest top-k probabilities. The top-$k$ generator takes time
$O(n(k+log(k))$, where $n$ is scan depth, or equivalently the number
of calls $Next()$.
\end{theorem}

\proofsketch
The correctness directly follows from the above discussions.

The time complexity of $O(n(k+log(k))$ does not take $Next()$ into
consideration. The initial of line 1-5 takes constant time. Each
call of $Prob()$ (Algorithm~\ref{alg:prob}) takes $O(k)$ time, based
on Theorem~\ref{the:prob}. Line 9, 11 take time $O(k)$. Line 10
takes time $O(log(k))$, due to the min-heap of size $k$. Line 6-11
are only executed $n$ times, so the total time complexity is
$O(n(k+log(k))$. \eop

\section{Experiment}

We have implemented our algorithm in Visual C++. We compare our
CondProb algorithm, denoted CP, for computing $p_{i,j}$, with the
heuristics proposed in \cite{SIGMOD08:Hua} which are RC (rule-tuple
compression only), RC+AR (RC with aggressive reordering), and RC+LR
(RC with lazy reordering).  The heuristics proposed can improve the
efficiency but they are algorithms in $O(kn^2)$, where $n$ is the
number of tuples and $k$ is the top-$k$ value.
The executable code and data generator used in \cite{SIGMOD08:Hua} are
downloadable\footnote{\url{http://www.cs.sfu.ca/~jpei/Software/PTKLib.rar}}.
We use exactly the same synthetic dataset as used
in~\cite{SIGMOD08:Hua}, which is also included in the package.

The parameters and default values are shown in \ctab
\ref{tab:para_tkp}.  Here, $mem$-$p$ is the expectation of the
membership probability of tuples, $p$ is the threshold specifying the
minimum top-$k$ probability of the result tuples returned, $k$ is the
top-$k$ value, $|rule|$ is the average number of tuples in a rule
($x$-tuple), $\#tuple$ is the total number of tuples, and $\#rule$ is
the total number of rules ($x$-tuples).

\begin{figure}[t]
\begin{center}
\begin{tabular}[t]{c}
    \subfigure[Vary $mem$-$p$]{
        \label{fig:exp:5a}
       \includegraphics[width=0.48\columnwidth,height=3.2cm]{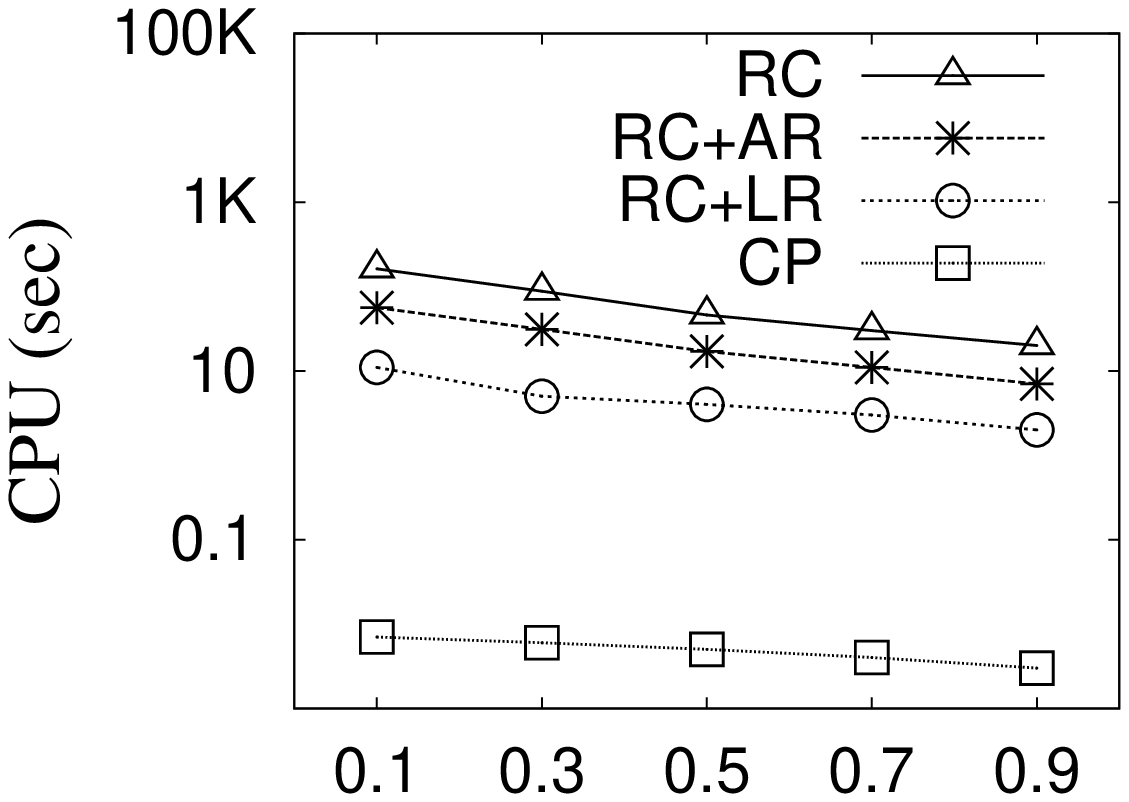}
    }
    \subfigure[Vary $|rule|$]{
        \label{fig:exp:5b}
       \includegraphics[width=0.48\columnwidth,height=3.2cm]{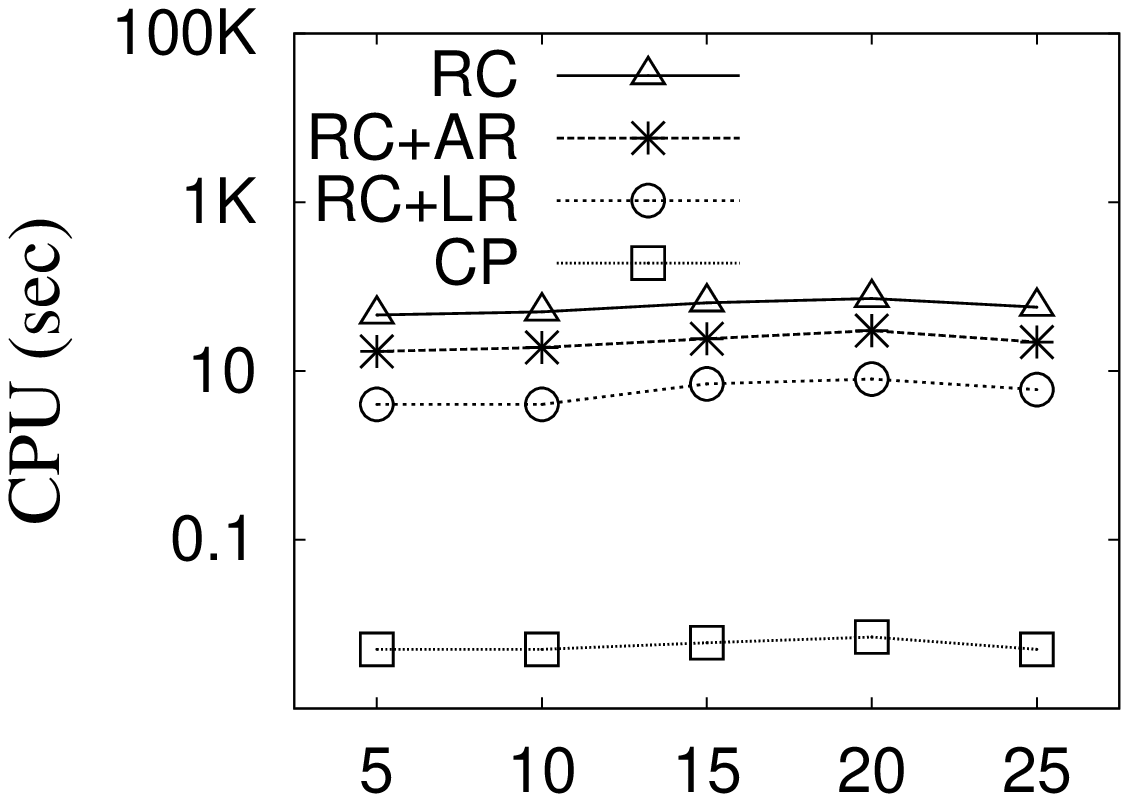}
    }
    \\
    \subfigure[Vary $k$]{
        \label{fig:exp:5c}
       \includegraphics[width=0.48\columnwidth,height=3.2cm]{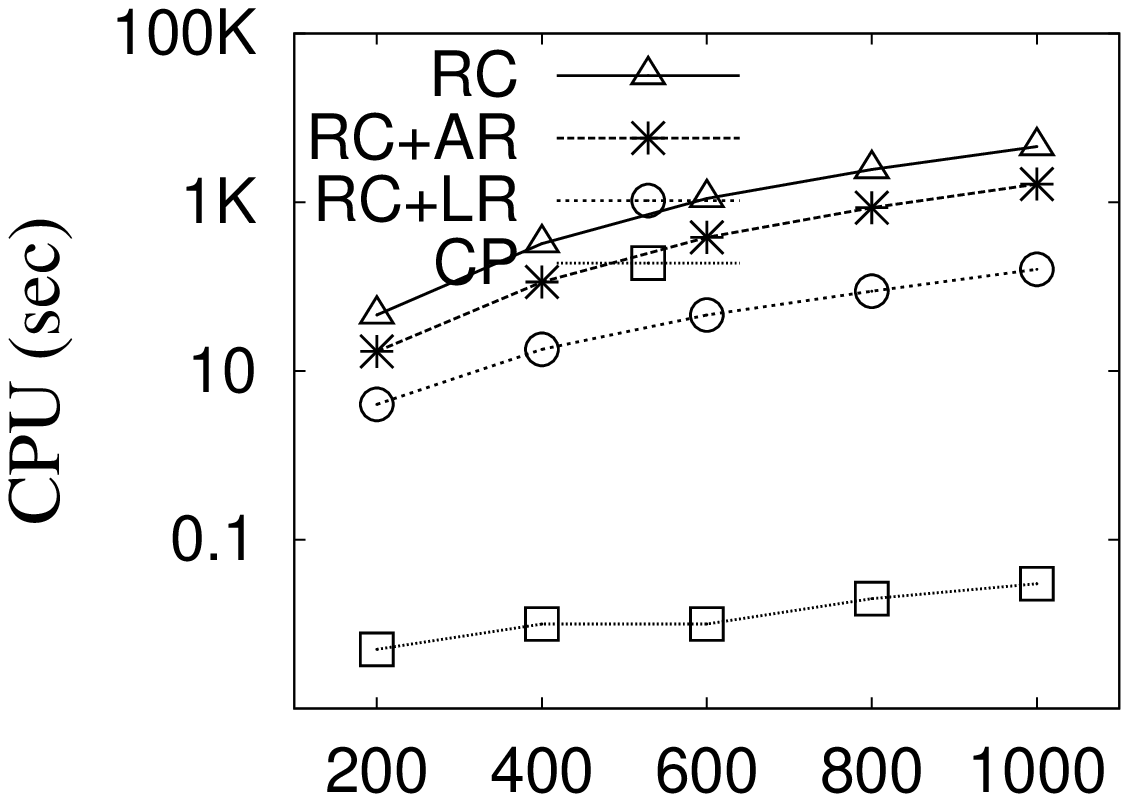}
    }
    \subfigure[Vary $p$]{
        \label{fig:exp:5d}
       \includegraphics[width=0.48\columnwidth,height=3.2cm]{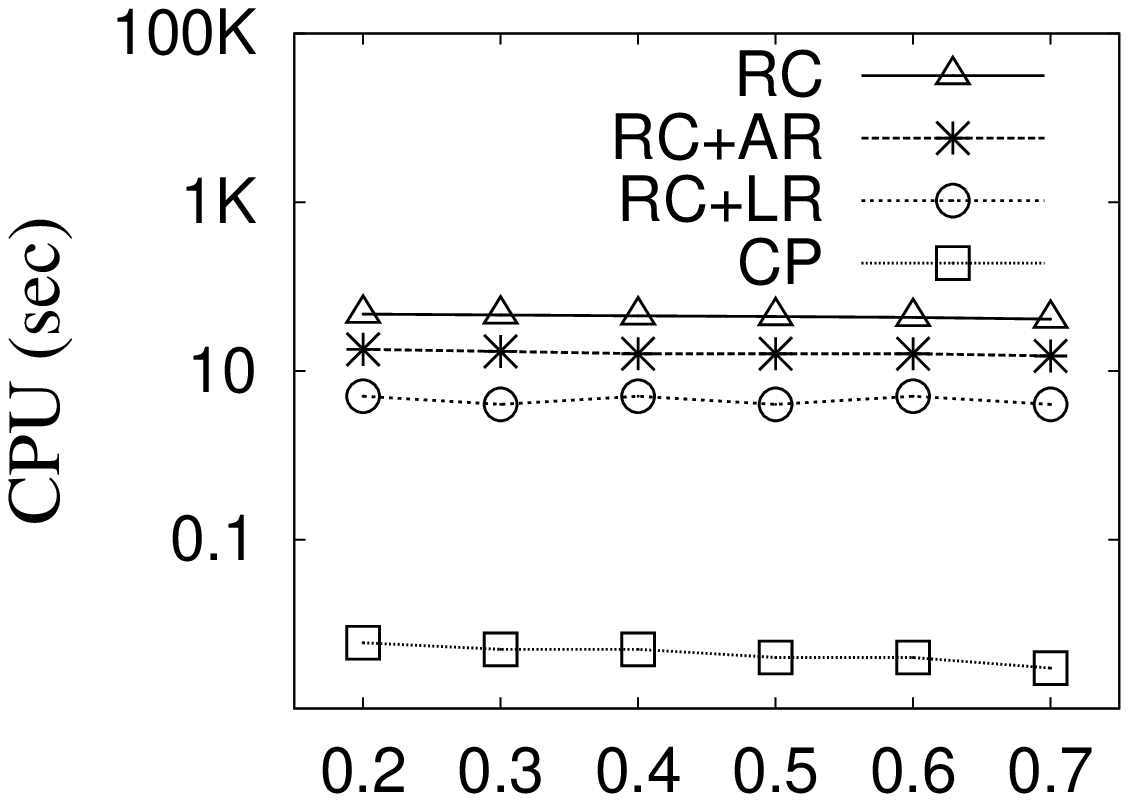}
    }
    \\
    \subfigure[Vary $\#tuple$]{
        \label{fig:exp:8a}
       \includegraphics[width=0.48\columnwidth,height=3.2cm]{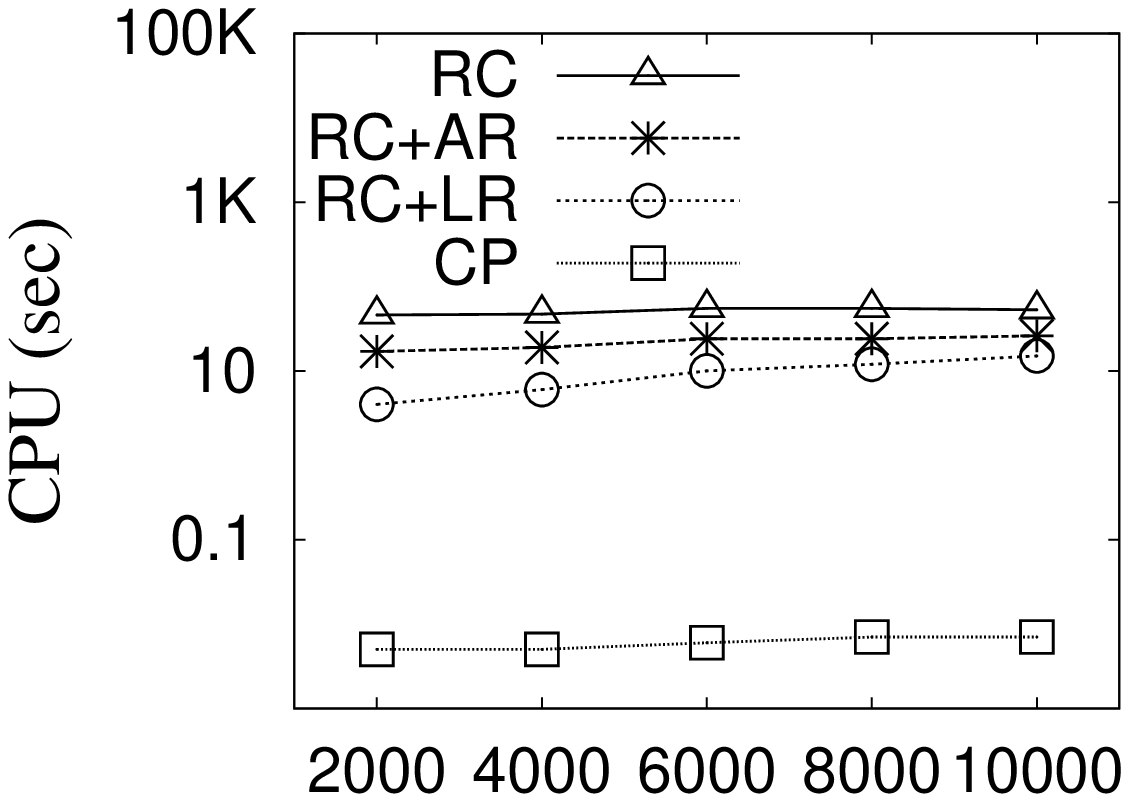}
    }
    \subfigure[Vary $\#rule$]{
        \label{fig:exp:8b}
       \includegraphics[width=0.48\columnwidth,height=3.2cm]{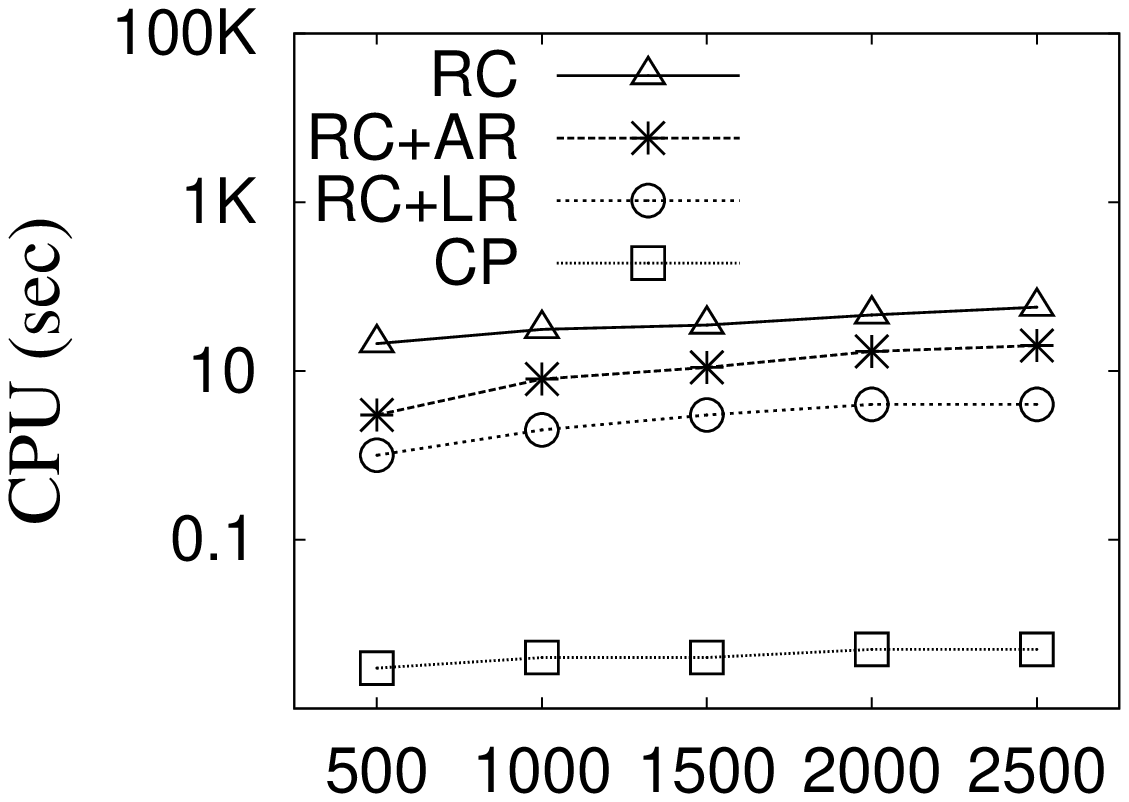}
    }
\end{tabular}
\end{center}
\vspace*{-0.4cm}
\caption{Computing $p_{i,j}$}
\label{fig:exp:tkp}
\end{figure}

The experimental results are shown in Fig.~\ref{fig:exp:tkp}. In all
figures, the shape of the curves for all the four algorithms are all
similar, our CP algorithm is $3,000$ times faster than RC+LR on
average, and $30,000$ times faster than RC on average.

\begin{table}[t]
\begin{center}
{
\begin{tabular}{|l|l|l|} \hline
\textbf{Parameter} & \textbf{Range} & \textbf{Default}\\
\hline \hline $mem$-$p$ & $0.1$, $0.3$, $0.5$, $0.7$, $0.9$ &
$0.5$\\
\hline $p$ & $0.1$, $0.3$, $0.5$, $0.7$, $0.9$ & $0.3$ \\
\hline $k$ & $200$, $400$, $600$, $800$, $1000$ & $200$ \\
\hline $|rule|$ & $5$, $10$, $15$, $20$, $25$ & $10$ \\
\hline $\#tuple$ &  $20000$, $40000$, $60000$, $80000$, $100000$ & $20000$\\
\hline $\#rule$ & $500$, $1000$, $1500$, $2000$, $2500$ & $2000$\\
\hline
\end{tabular}
}
\end{center} \vspace*{-0.4cm}
\caption{Parameters and Default Values}
\label{tab:para_tkp}
\end{table}

\section{Related work}
\label{sec:related}

Uncertain data has received increasing attention recently, most of
them represent the uncertainty as probability values, also called
probabilistic data. Many probabilistic data model and systems have
been proposed, for example, Trio system~\cite{VLDB06:Agrawal},
MystiQ system~\cite{VLDBJ07:Dalvi}, MayBMS
system~\cite{ICDE08:Antova}.

In the literature, several works study computing the top-$k$ results
by the interplay of score and probability, based on the possible
worlds semantic.
%
%
U-Topk and U-kRanks queries are first proposed in
\cite{ICDE07:Soliman} on a general uncertain data model.
\cite{ICDE08:Yi,TKDE08:Yi} improve the performance of the U-Topk and
U-kRanks queries using a dynamic programming approach, under an
$x$-Relation model, by utilizing the independent and mutually
exclusive relationship between tuples.
\cite{ICDE08:Hua,SIGMOD08:Hua} define the PT-k query, and propose
three heuristic approaches to answer the PT-k queries.
In~\cite{ICDE08:Yi,TKDE08:Yi,ICDE08:Hua,SIGMOD08:Hua}, to answer a
U-kRanks or PT-k query, they all need to compute $p_{i,j}$, the
probability that tuple $t_i$ ranks at the $j$-th position in possible
worlds, for $1 \leq i \leq n$ and $1 \leq j \leq k$, with the time
complexity $O(kn^2)$.
\cite{VLDB08:Jin} adapt the U-Topk/U-kRanks/Global-Topk (Global-Topk
\cite{DBRank08:Zhang} is the same as Pk-topk in~\cite{VLDB08:Jin})
queries in a uncertain stream environment under a sliding-window
model, and design both space- and time-efficient synopses to
continuously monitor the top-$k$ results.  But, \cite{VLDB08:Jin} only
consider the single-alternative case, or in other words, all tuples
are independent.
\cite{SSDBM08:Bernecker,EDBT08:Lian} also need to compute the
$p_{i,j}$ values, running the probabilistic ranking in a middleware to
answer ranking spatial queries on uncertain spatial data.
\cite{TODS08:Soliman} discusses aggregate queries.

There are also works that find the top-$k$ results based on the
probability only.
\comment{ Ljosa et al.~\cite{ICDE08:Ljosa} find the top-$k$ spatial
joins of probabilistic objects, where each point has a probability
to belong to a specific object, these probabilities are integrated
into the score function, ranking is based on the score function. }
In~\cite{ICDE07:Re}, Re et al. find the $k$ most probable answers
for a given general SQL query. In this scenario, each answer has a
probability instead of a score, which intuitively represents the
confidence of its existence, ranking is only based on probabilities.
They use Monte Carlo simulations to get the top-$k$ results
efficiently, as in general it is \#P-complete to get the existence
probability~\cite{VLDBJ07:Dalvi}.
\cite{VLDB07:Pei,SIGMOD08:Lian,VLDB08:Beskales} retrieve $k$
objects from a uncertain spatial database, that have the highest
probability to be a skyline point or nearest neighbor.

\section{Conclusion}

The probabilistic top-$k$ queries based on the interplay of score
and probability, under the possible worlds semantic, become an
important research issue that considers both score and uncertainty
on the same basis.  In the literature, many different probabilistic
top-$k$ queries are proposed. In the $x$-Relational model, an
$x$-tuple consists of a set of mutually exclusive tuples to
represent a discrete probability distribution of the possible tuples
in a randomly instantiated data.  Almost all of them need to compute
the probability of a tuple $t_i$ to be ranked at the $j$-th position
across the entire set of possible worlds. We call it $p_{i,j}$ computing.
The cost of computing $p_{i,j}$ is the dominant cost and is known as
$O(kn^2)$, where $n$ is the size of dataset. In this paper, we
proposed a new novel algorithm that computes such probability
efficiently based on conditional probability and the system of
linear equations. We proved the correctness of our approach, and
showed that the time complexity is $O(kn)$. We confirmed the
efficiency by comparing our approach with the up-to-date heuristics
and found that our approach can be at least $3,000$ times faster.

\bibliographystyle{abbrv}
\bibliography{sigproc}

\end{document}